\documentclass[final,1p,times]{elsarticle} 

\usepackage{graphicx}
\usepackage{amssymb} 
\usepackage{amsthm} 
\usepackage{lineno}

\pdfmapfile{+txfonts.map}

\newcommand{\be}{\begin{equation}}
\newcommand{\ee}{\end{equation}}
\newcommand{\ba}{\begin{eqnarray}}
\newcommand{\ea}{\end{eqnarray}}
\newcommand{\bd}{\begin{displaymath}}
\newcommand{\ed}{\end{displaymath}}

\def\oneth{{\textstyle{\frac{1}{3}}}}

\journal{Nuclear Physics A} 

\begin{document}

\begin{frontmatter} 

\title{Causal Baryon Diffusion and Colored Noise for Heavy Ion Collisions}

\author{J. I. Kapusta$^1$ and C. Young}
\fntext[col1] {Presenter.}
\address{School of Physics \& Astronomy, University of Minnesota, Minneapolis, MN 55455, USA}


\begin{abstract}
 
We construct a model for baryon diffusion which has the desired properties of analyticity and causality.  The model also has the desired property that the noise correlation function is not a Dirac delta function in space and time.  The model depends on three time constants in addition to the diffusion constant.  This description can be incorporated into 2nd order viscous hydrodynamical models of heavy ion collisions. 

\end{abstract} 

\end{frontmatter} 



\section{Introduction}

There are at least four sources of fluctuations in high energy heavy ion collisions.  These include initial state fluctuations, hydrodynamic fluctuations (noise) due to finite particle number effects in coarse grained fluid elements, energy and momentum deposition by jets traversing the medium, and freeze-out fluctuations as fluid elements are converted to individual particles.  The focus here is on hydrodynamic fluctuations.  The fluctuation-dissipation theorem says that any dissipative system will necessarily have noise \cite{Landau:1980st}.  Since the matter produced in high energy heavy ion collisions can be described very well by second order dissipative fluid dynamics, it behooves us to investigate the associated fluctuations, especially as their measurement might provide valuable information on the transport coefficients of matter at high energy density.  The first exploratory studies along these lines were carried out in the context of the one dimensional hydrodynamic model of Bjorken \cite{Kapusta:2011gt,JoeJuan}.  Here we consider the general problem of baryon number diffusion, fluctuation, and noise.  For simplicity we consider a system at constant temperature with small variations in baryon density or chemical potential.

In the Landau-Lifshitz definition of the local rest frame the baryon current is expressed as $J^{\mu} = n u^{\mu} + \Delta J^{\mu}$.  Here $n$ is the proper local baryon density and $\Delta J^{\mu}$ is the dissipative part.  The dissipative part must satisfy $u_{\mu} \Delta J^{\mu} = 0$ so that $n$ represents the proper baryon density.  In first order viscous fluid dynamics it takes the form $\Delta J^{\mu} = \sigma T \Delta^{\mu} \left(\beta \mu \right)$ where $\beta = 1/T$, $\mu$ is the chemical potential, $\sigma = D (\partial n/\partial \mu)$ is the baryon conductivity with $D$ the diffusion constant, and $\Delta_{\mu} = \partial_{\mu} - u_{\mu} \left( u \cdot \partial \right)$ is a derivative normal to $u^{\mu}$.  For baryon diffusion in a system with no energy flow one obtains the usual diffusion equation
As is well known \cite{Joseph,Aziz}, the diffusion equation results in instantaneous transport and is not suitable for numerical hydrodynamic simulations of high energy heavy ion collisions.
Adding a noise term $I^{\mu}$ to the current results in the noise correlator
\be
\langle I^i I^j (t, {\bf x}) \rangle = 2 \sigma T  \delta ({\bf x}) \, \delta(t) \, \delta_{ij} \,.
\label{Dnoise}
\ee 
This is white noise (independent of frequency and wave number when Fourier transformed).  It is can be used satisfactorily when noise is treated perturbatively as in \cite{Kapusta:2011gt,JoeJuan}, but it creates severe problems when noise is treated nonperturbatively with small grid sizes in numerical fluid dynamics \cite{Young:2013fka}.

In what follows we will discuss how the diffusion equation can be modified, and what the corresponding dissipative part of the current ought to be, in order that baryon number is transported with a finite speed and the noise correlator has a finite range in space and time.

\section{Generalizations of the Diffusion Equation and Associated Baryon Current}

The problem we are discussing is related to the problem of heat conduction.  The difference is that heat conduction is associated with the dissipative part of the energy-momentum tensor, whereas baryon diffusion is associated with the dissipative part of the conserved current.  Consider, therefore, the Gurtin-Pipkin equation \cite{GP,Joseph}
\be
\left[ \frac{\partial}{\partial t} - D\nabla^2 + \tau_1 \frac{\partial^2}{\partial t^2} + \tau_2^2 \frac{\partial^3}{\partial t^3}
- \tau_3' D\frac{\partial}{\partial t}\nabla^2  \right] n = 0 \,.
\ee
(The reason for the prime will soon become apparent.)  This equation is hyperbolic and transports information with a finite speed.  When one sets $\tau_2 = \tau_3' = 0$ one obtains the Cattaneo equation \cite{Cattaneo,Joseph} which is also hyperbolic.  When one also sets $\tau_1 = 0$ one obtains the usual diffusion equation, which is parabolic.  The Gurtin=Pipkin equation follows from the dissipative current
\be
\Delta J^\mu = D \Delta^{\mu} \frac{1 + \tau_4(u \cdot \partial)}{1+\tau_1(u \cdot \partial) + \tau_2^2(u \cdot \partial)^2
 + \tau_3 D \Delta^2} \, n
\ee
where the differential operator in the denominator is to be understood as its Taylor series expansion, and $\tau_3' = \tau_3 + \tau_4$. 

The baryon density response function is
\be
G_R(\omega, {\bf k}) = \left(\frac{\partial n}{\partial \mu} \right) \frac{\omega}{A(\omega, {\bf k})}
\ee
where
\be
A(\omega, {\bf k}) \equiv \omega + \frac{iDk^2(1-i\tau_4 \omega)}{1-i\tau_1\omega - \tau_2^2 \omega^2+ \tau_3 D k^2} \, .
\label{A}
\ee
The fluctuation-dissipation theorem can be employed to calculate the density correlation function
\be
\left\langle \delta n \delta n(\omega, {\bf k}) \right \rangle = -\frac{2T}{\omega}{\rm Im}\left\{ G_R \right\} 
= 2 T \left(\frac{\partial n}{\partial \mu} \right) \frac{{\rm Im}\left\{ A \right\}}{|A|^2}
= i T \left(\frac{\partial n}{\partial \mu} \right) \left( \frac{1}{A} - \frac{1}{A^*} \right)
\ee
and the noise correlation function
\be
\oneth k^2 \langle I^l I^l ({\bf k},\omega) \rangle =  
A(\omega, {\bf k}) A^*(\omega, {\bf k}) \langle \delta n \delta n({\bf k},\omega \rangle =
- i T \left(\frac{\partial n}{\partial \mu} \right) \left( A - A^* \right)
\label{gen_noise_cor}
\ee
Interestingly, the poles of $A$ determine the behavior of the density correlator but the zeroes determine the behavior of the noise correlator.

\section{Results}

The density correlation function for the normal diffusion equation has the familiar form
\be
\left\langle \delta n \delta n(t, {\bf x}) \right \rangle = T \left(\frac{\partial n}{\partial \mu} \right) 
\left(\frac{1}{4 \pi D t}\right)^{3/2} {\rm e}^{-r^2/4 D t} \,,
\ee
so that baryon diffusion happens with infinite speed of propagation.  The noise correlator was given in the introduction.

\begin{figure}[h]
\begin{center}
\includegraphics*[width=7.00cm]{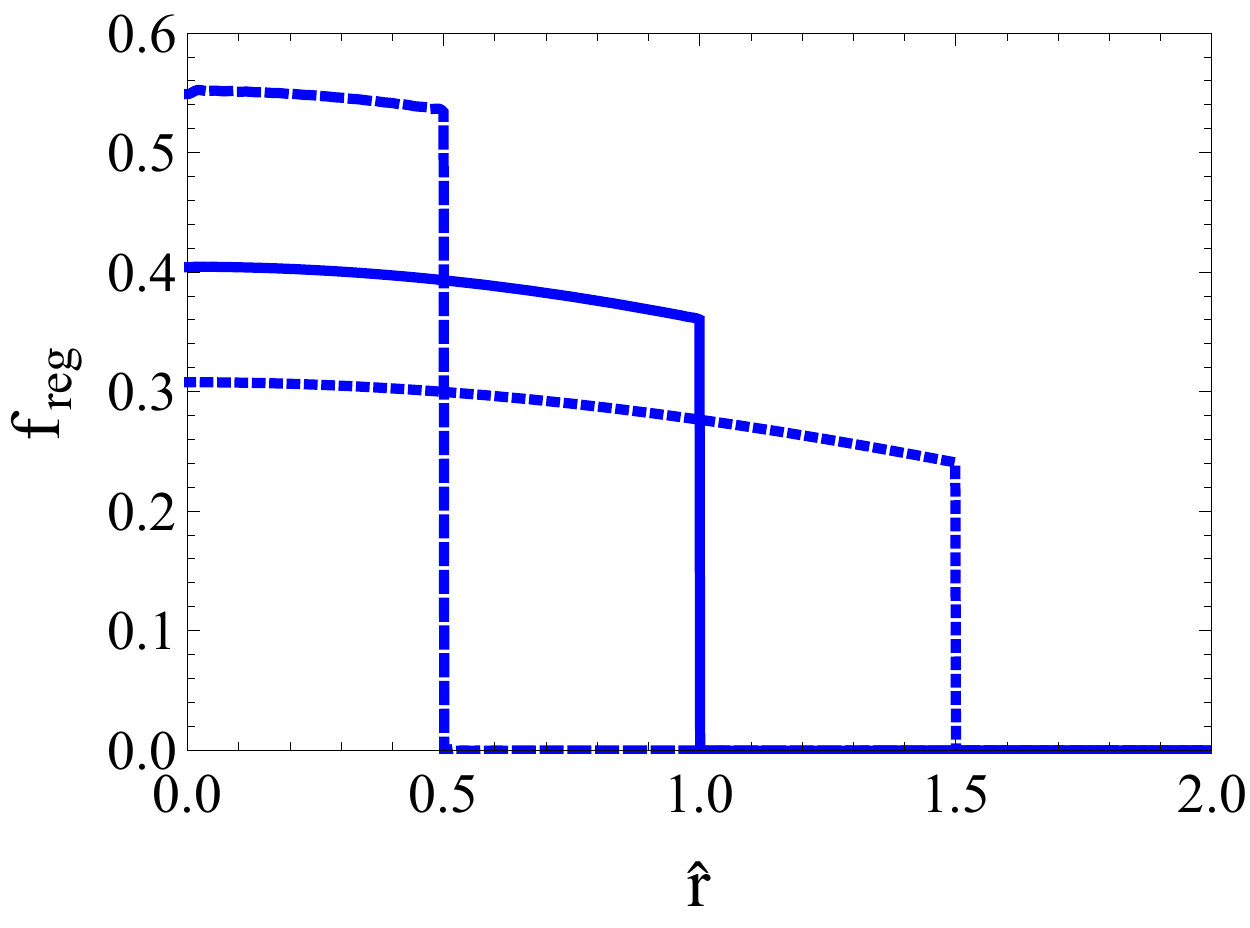}
\caption{(color online)  The regular part of the dimensionless density correlator $f_{\rm reg}(\hat{r},\hat{t})$ for the Cattaneo equation.
  The correlator is shown for $\hat{t}=0.5$ (dashed line),  $\hat{t}=1$ (solid line), and  $\hat{t}=1.5$ (dotted line).
From Ref. \cite{JoeClint}.}
\label{Fig1}
\end{center}
\end{figure}

The density correlator for the Cattaneo equation has a pair of imaginary poles for $k < k_c$ and a pair of complex poles for $k > k_c$, where $k_c = 1/4 \tau_1 D$.  They always lie in the lower half plane so that causality is respected.  The group velocity is $v_g = v_0 k / \sqrt{k^2 - k_c^2}$ where $v_0 = \sqrt{D/\tau_1}$.  The fact that it diverges is not a problem since near $k_c$ wave packets are severely distorted and $v_g$ does not represent the speed of the center of a wavepacket \cite{Brillouin}.  Apart from some unimportant prefactors the dimensionless correlator can be written as
\be
f(\hat{r},\hat{t}) =
\frac{\pi}{2} \frac{{\rm e}^{-\hat{t}}}{\hat{r}} \left[ \left(1 + \frac{\hat{t}}{2}\right) \delta (\hat{r}-\hat{t}) - \delta' (\hat{r}-\hat{t})
 -  \frac{(4+\hat{t})\hat{t}}{8} \theta (\hat{r}-\hat{t})  \right]
+ f_{\rm reg}(\hat{r},\hat{t})
\ee
where $\hat{r}=r/2v_0 \tau_1$ and $\hat{t} = t/2\tau_1$.  The appearance of the Dirac delta function and its derivative, followed by a diffusion wake, is very similar to what was found for a different response function in \cite{Kapusta:2011gt}.  The regular part is shown in Figure \ref{Fig1}; the speed of propagation is clearly 1 in these dimensionless units ($v_0$ in dimensional units).  The noise correlator is 
\be
\langle I^i I^j (t, {\bf x}) \rangle = \frac{\sigma T}{\tau_1} \delta ({\bf x}) \, {\rm e}^{-|t|/\tau_1} \, \delta_{ij}
\label{Cnoise}
\ee
so that it has a finite range in time but is still a delta function in space.

\begin{figure}[h] 
\begin{center}
      \includegraphics[width=6.6cm]{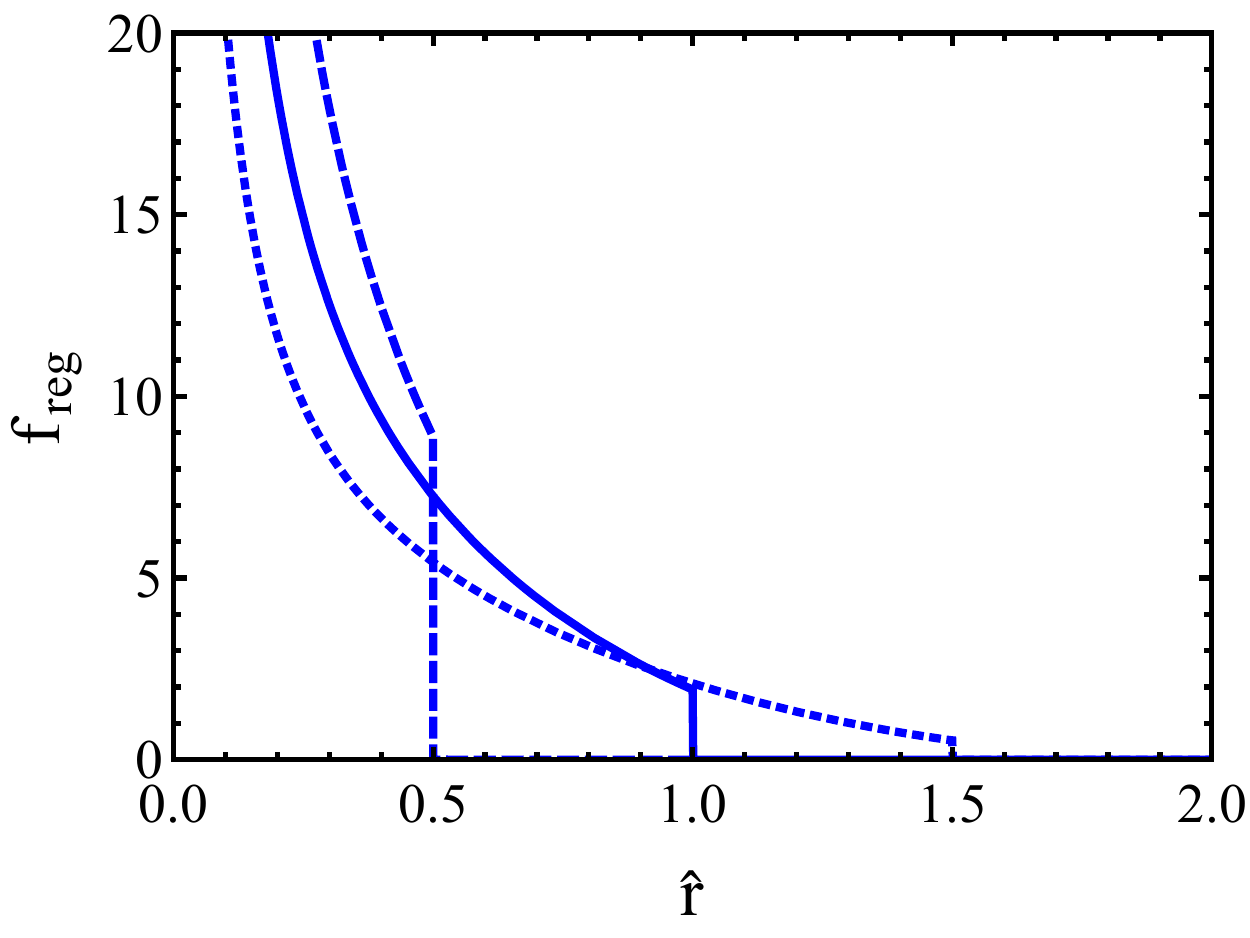}
            \includegraphics[width=6.6cm]{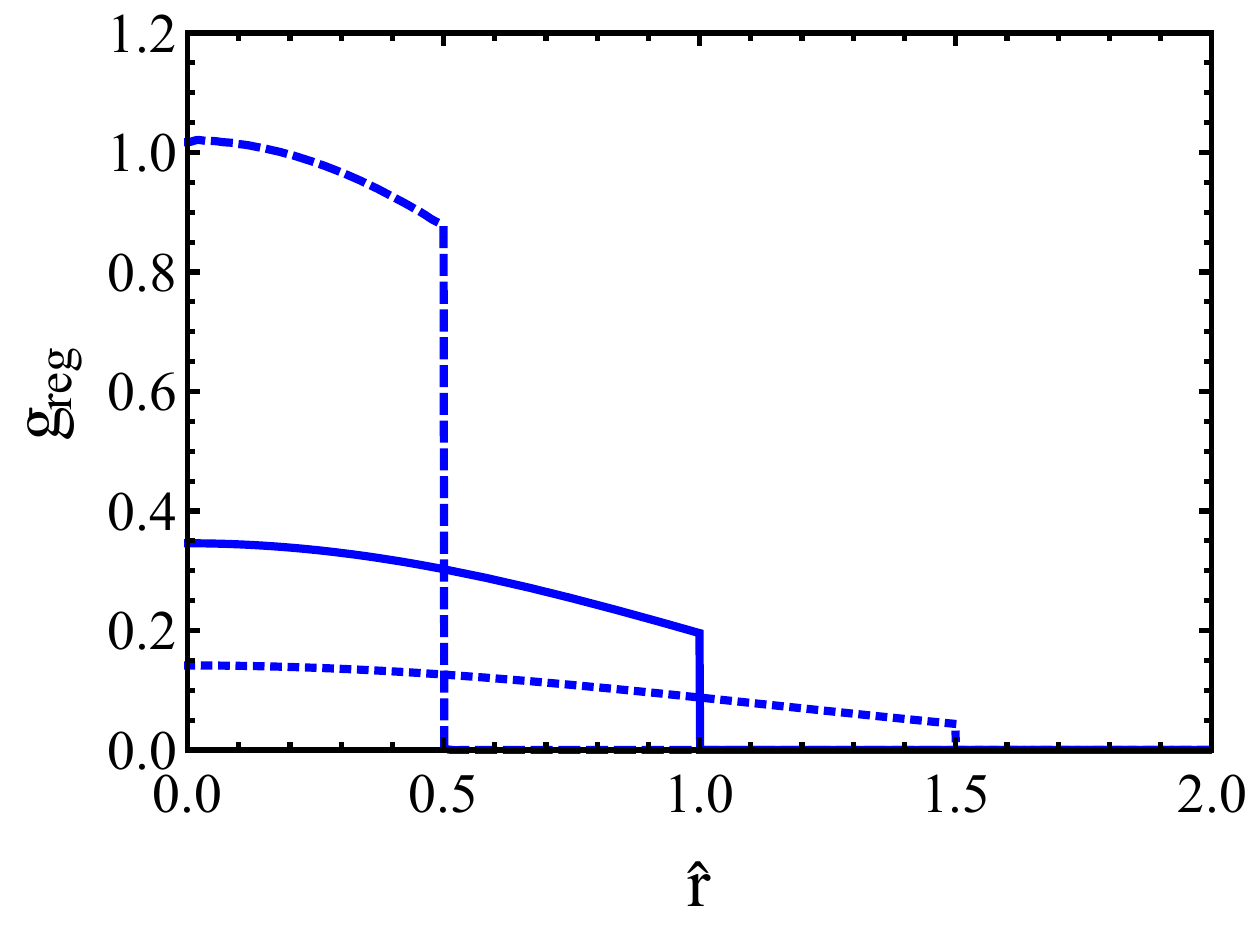}
 \end{center}
\caption{(color online)  The left panel shows the regular part of the dimensionless density correlator $f_{\rm reg}(\hat{r},\hat{t})$ for the Gurtin-Pipkin equation for $\hat{t}=0.5$ (dashed line),  $\hat{t}=1$ (solid line), and  $\hat{t}=1.5$ (dotted line).  The right panel shows the regular part of the dimensionless noise correlator 
$g_{\rm reg}(\hat{r},\hat{t})$.  From Ref. \cite{JoeClint}.}
\label{Fig2}
\end{figure}

The density correlator for the Gurtin-Pipkin equation has 3 poles which are all in the lower half plane if $\tau_2^2 < \tau_1 \tau_3'$.  The asymptotic group velocity is $v_0 = \sqrt{\tau_3' D/\tau_2^2}$.  The dimensionless form of the correlator is
\be
f(\hat{r},\hat{t}) = A \delta({\bf x}) \exp(-a \hat{t}) + B \frac{\exp(-b \hat{t})}{\hat{r}} \delta(\hat{r}-\hat{t}) +  f_{\rm reg}(\hat{r},\hat{t})
\ee
once again displaying a singular part and a regular part.  An example of the latter is plotted in the left panel of Figure \ref{Fig2} using $\tau_1 = 3\tau_2$, $\tau_4 = 0$ and $v_0^2 = 1/3$.

When $\tau_1 > 2\tau_2$, the noise correlator has a pair of imaginary poles for $k < k_c$ and a pair of complex poles for $k > k_c$, where now $k_c^2 = (\tau_1^2/\tau_2^2 - 4)/4 \tau_3 D$.  They always lie in the lower half plane so that causality is respected.  The group velocity is $v_g = v_0 k / \sqrt{k^2 - k_c^2}$ where $v_0$ is the same as above if $\tau_4 = 0$.  When $\tau_1 < 2\tau_2$, there are a pair of complex poles, and the group velocity takes the form $v_g = v_0 k / \sqrt{k^2 + k_0^2}$.  The functional form of the dimensionless noise correlator is
\be
g(\hat{r},\hat{t}) = \frac{\pi}{2} \frac{\exp(-\tau_1 \hat{t}/\tau_2)}{\hat{r}} \delta(\hat{r}-\hat{t}) +  g_{\rm reg}(\hat{r},\hat{t}) \,.
\ee    
An example is plotted in the right panel of Figure \ref{Fig2} using the same parameters as above.  This obviously is colored noise, both in space and in time.  An interesting footnote is that there is no wake behind the front when $\tau_1 = 2 \tau_2$.

\section{Conclusions}

We have studied and compared the baryon current in 1st, 2nd and 3rd order dissipative fluid dynamics.  With no energy transport but only pure baryon diffusion, these correspond to the ordinary heat conduction equation, the Cattaneo heat conduction equation, and the Gurtin-Pipkin heat conduction equation.  Using the fluctuation-dissipation theorem we computed the response function, the baryon density correlation function, and the noise correlation function.  One needs at least 2nd order for finite propagation speed, and at least 3rd order for finite correlation lengths and times for noise.  More details of this work may be found in \cite{JoeClint}.

In the future these results could be readily implemented in numerical hydrodynamic codes.  Baryon transport and noise should be important even if the net baryon number is zero, and may play a crucial role near a critical point \cite{JoeJuan}.  Finally, microscopic calculations are needed to compute the time constants, which in reality are probably functions of temperature and density.  We look forward to progress in these and related avenues of investigation.  Noise happens!

This work was supported by the U.S. DOE Grant No. DE-FG02-87ER40328.








\end{document}